\documentclass[preprint,showpacs,preprintnumbers,amsmath,amssymb,natbib]{revtex4}

\usepackage{graphicx}% Include figure files
\usepackage{epsfig}		
\usepackage{dcolumn}% Align table columns on decimal point
\usepackage{bm}% bold math
\def\0{\mbox{\tiny $0$}}
\def\1{\mbox{\tiny $1$}}
\def\2{\mbox{\tiny $2$}}
\def\3{\mbox{\tiny $3$}}
\def\4{\mbox{\tiny $4$}}
\def\5{\mbox{\tiny $5$}}
\def\6{\mbox{\tiny $6$}}
\def\7{\mbox{\tiny $7$}}
\def\8{\mbox{\tiny $8$}}
\def\9{\mbox{\tiny $9$}}

\def\f14{\mbox{\tiny $\frac{1}{4}$}}

\def\bb#1{\mbox{\footnotesize $(#1)$}}
\begin{document}
%
%%%%%%%%%%%%%%%%%%%%%%%%%%%%%%%% PAPER %%%%%%%%%%%%%%%%%%%%%%%%%%%%%%%%%%%%%

\title{The C$\nu$B energy density through the quantum measurement theory}
\author{A. E. Bernardini}
\email{alexeb@ufscar.br}
\author{V. A. S. V. Bittencourt}
\affiliation{Departamento de F\'{\i}sica, Universidade Federal de S\~ao Carlos, PO Box 676, 13565-905, S\~ao Carlos, SP, Brasil}

\date{\today}

\begin{abstract}
We apply concepts from the quantum measurement theory to obtain some cosmological neutrino background (C$\nu$B) properties and discuss their relevance in defining theoretical bounds on cosmological neutrino energy density.
Describing three neutrino generations as a composite quantum system through the generalized theory of quantum measurement provides us with the probabilistic correlation between observable energies and neutrino flavor eigenstates.
By observing that flavor-{\em averaged} and flavor-{\em weighted} energies are the quantum observables respectively generated by {\em selective} and {\em non-selective} quantum measurement schemes, it is possible to identify the constraints on the effective mass value expression that determines the neutrino contribution to the energy density of the cosmic inventory.
Our results agree with the quantum mechanics viewpoint that asserts that the cosmological neutrino energy density is obtained from a coherent sum of mass eigenstate energies, for normal and inverted mass hierarchies.
\end{abstract}



\maketitle

\newpage
The cosmological neutrino background represents a relevant fraction of the cosmic inventory and, therefore, has a perpetual influence on the cosmological evolution of the background Universe and on the propagation of linear perturbations \cite{Dolgov02,Dodelson,Boy10}.
As an example, when one quantifies the matter power spectrum of the Universe, the fraction of cold and/or hot dark matter corresponding to neutrinos depends on the value of the neutrino rest mass that indeed defines whether its contribution is relevant to the formation of large scale structures.
In this context, some recent issues \cite{Fuller,Kis08,Bel99,Dol02} on quantum mechanics of cosmological neutrinos have focused on finding the most adequate procedure for extracting neutrino mass values from cosmological data.

Cosmology is at first order sensitive to the total neutrino mass if all states have the same number density (i. e. if the cosmological neutrino flavor ensemble is a maximal statistical mixing), providing information on the absolute value of the mass but blind to neutrino mixing angles or possible CP violating phases \cite{Pas06}.
Such cosmological results are complementary to terrestrial experiments as beta decay and neutrinoless double beta decay, which are respectively sensitive to the following formulations for the neutrino effective mass,
\begin{equation}
m_{\beta} = \left(\sum_{i} |U_{ei}|^{2} m_{i}^{2}\right)^{\frac{1}{2}},
\label{1}
\end{equation}
that corresponds to an {\em averaged} mass, and
\begin{equation}
m_{\beta\beta} = \left|\sum_{i} U_{ei}^{2} m_{i}\right|
\end{equation}
which corresponds to a {\em weighted} mass, where both are defined through the peculiarities of the interactions at detection procedures and their relations to the measurement techniques \cite{Fassler}.

Turning to the context of cosmological neutrinos, we shall discuss the theoretical derivation of the effective mass expressions through the framework of the quantum measurement theory.
Assuming that three neutrino generations can be described by a statistical mixture of flavor eigenstates, flavor-{\em weighted} energies will be introduced in order to set the energy properties of composite quantum systems probabilistically correlated to flavor quantum numbers (see the Appendix I).
The corresponding theoretical background used for defining flavor{\em averaged} and flavor-{\em weighted} energies is provided by the generalized theory of quantum measurement, where such energy definitions are respectively identified with {\em selective} and {\em non-selective} quantum measurement schemes \cite{Breuer}.
In particular, we shall focus our attention on how the definition of flavor associated energies can affect the theoretical mass predictions for cosmological neutrinos.

The usual single-particle quantum definition of flavor-{\em averaged} energies is ambiguous \cite{Bernardini} since, from the fundamentals of the quantum oscillation phenomena, an arbitrary $\alpha$-flavor eigenstate can be partially, or even completely, converted into a $\beta$-flavor eigenstate, with $\alpha \neq \beta$.
Assuming that the time evolution of the system is driven by a diagonal Hamiltonian in the mass eigenstate basis, one can notice that flavor energy ``measurements'' or ``projections'' sometimes correspond to crude definitions.
Extracting effective values of neutrino masses from measurable quantum observables can then become an ambiguous procedure.
To avoid ambiguities and misunderstandings in the interpretation of how the energies can be associated to flavor eigenstates and correlated to flavor probability measurements, we consider some principles of the generalized measurement theory.

Once the Hamiltonian of a neutrino system in the three mass eigenstate basis is diagonal, $H = Diag\{E_{\1},\, E_{\2},\,E_{\3}\}$, the $\alpha$-flavor projection operators can be easily defined as \cite{Giu91,Zra98,Zub98,Giu98}
\begin{equation}
M^{\alpha}_{(t)} = |\nu^{\alpha}_{(t)} \rangle \langle \nu^{\alpha}_{(t)}|,
\label{eq01B}
\end{equation}
where $\alpha = e,\,\mu,\,\tau$, and $|\nu^{\alpha}_{(0)}\rangle = \sum_s U^{*}_{\alpha s}\,|\nu^s_{(0)}\rangle$, with $s = 1,\,2,\,3$ denoting the indices for mass eigenstates, with corresponding vacuum mass eigenvalues $m_s$.
The unitary transformation matrix elements $U^{*}_{\alpha s}$ are parameterized by three mixing angles, $\theta_{12},\, \theta_{23},\,\theta_{13}$, where the CP-violating phase $\delta$ is omitted at this first analysis.

Considering that the density matrix representation of a composite quantum system of three flavor species is given by
\begin{equation}
\hat{\rho}\bb{t} \equiv \hat{\rho} = \sum_{\alpha = e,\mu,\tau}w_{\alpha} \, M^{\alpha}_{(t)} ~~~\mbox{with} ~~ \sum_{\alpha = e,\mu,\tau} w_{\alpha} = 1 ~~\mbox{and} ~~ \sum_{\alpha = e,\mu,\tau} M^{\alpha}_{(t)}  = \mathbf{1},
\label{eq05}
\end{equation}
one easily finds that, from the unitarity expressed above, the re-defined probabilities of measuring $\alpha$-flavor eigenstates at time $t$ are given by
\begin{eqnarray}
P^{\alpha}_{(t)}   = \mbox{Tr}\{M^{\alpha}_{(0)}\, \hat{\rho}\} &=& \sum_{\beta = e,\mu,\tau}{w_{\beta} \, \mbox{Tr}\{M^{\alpha}_{(0)}\,M^{\beta}_{(t)}\}} = %\nonumber\\&=&
\sum_{\beta = e,\mu,\tau}{w_{\beta} \,\mathcal{P}_{\alpha\rightarrow \beta}\bb{t}},
\label{eq06}
\end{eqnarray}
where $\mathcal{P}_{\alpha\rightarrow \beta}\bb{t} =  |\langle \nu^{\beta}_{(0)}\mid\nu^{\alpha}_{(t)} \rangle|^{\2}$ describe the oscillation probabilities in the single-particle quantum mechanics framework.
Since the unitarity is expressed by
\begin{eqnarray}
\sum_{\alpha = e,\mu,\tau}P^{\alpha}_{(t)}
                        &=& \sum_{\alpha,\beta = e,\mu,\tau}{w_{\beta}\, \mathcal{P}_{\alpha\rightarrow \beta}\bb{t}} =   \sum_{\beta = e,\mu,\tau}{w_{\beta} \left(\sum_{\alpha = e,\mu,\tau}\mathcal{P}_{\alpha\rightarrow \beta}\bb{t}\right)} = 
\sum_{\beta = e,\mu,\tau}{w_{\beta}} = 1,
\label{eq06B}
\end{eqnarray}
we shall verify that the definition from Eq.~(\ref{eq06}) allows one re-interpret the energies related to each flavor quantum number.

The generalized theory of quantum measurements \cite{Breuer,10,20,30} is based on the extended idea of a {\em positive operator-valued} measure which associates with each measurement outcome $\alpha$ a positive operator $M^{\alpha}_{(0)}$.
It can be interpreted in terms of the von-Neumann-L\"{u}ders projection postulate that introduces the concepts of {\em selective} and {\em non-selective} measurements \cite{Breuer}.
The measurement outcome $\alpha$ represents a classical random number with probability distribution given by Eq.~(\ref{eq06}) where $M^{\alpha}_{(0)}$ is a positive operator called the {\em effect}.
For the case that the measurement is a {\em selective} one \cite{Breuer}, the sub-ensemble of those systems for which the outcome $\alpha$ has been found has to be described by the density matrix
\begin{equation}
\hat{\rho}_{\alpha} = \left(P^{\alpha}_{(t)}\right)^{-1} \, M^{\alpha}_{(0)}\, \hat{\rho} \, M^{\alpha}_{(0)}
\label{eq14},
\end{equation}
where $M^{\alpha}_{(0)}\, \hat{\rho} \, M^{\alpha}_{(0)}$ is called {\em operation}, which maps positive into positive operators.
Notice that one consistently has
\begin{equation}
\mbox{Tr}\{\hat{\rho}_{\alpha}\} = \left(P^{\alpha}_{(t)}\right)^{-1} \,  \mbox{Tr}\{ M^{\alpha}_{(0)}\, \hat{\rho} \, M^{\alpha}_{(0)}\}
= \left(P^{\alpha}_{(t)}\right)^{-1} \,  \mbox{Tr}\{ M^{\alpha}_{(0)}\, \hat{\rho}\} = 1.
\label{eq14C}
\end{equation}
For the corresponding {\em non-selective} measurement \cite{Breuer} one has the density matrix
\begin{equation}
\hat{\rho}^{\prime} = \sum_{\alpha} {P^{\alpha}_{(t)} \hat{\rho}_{\alpha}},
\label{eq15}
\end{equation}
from which it is also easily verified that $\mbox{Tr}\{\hat{\rho}^{\prime}\} = 1$.

By extending the theoretical constructions described in the Appendix I to three flavor generations, the flavor-{\em averaged} energies, $E^{\alpha}_{(t)}$, are therefore computed through the density matrix for {\em selective} measurements, $\hat{\rho}_{\alpha}$, as \cite{Bernardini}
\begin{equation}
E^{\alpha}_{(t)} = \mbox{Tr}\{H \, \hat{\rho}_{\alpha}\}
\label{eq14B},
\end{equation}
and the sum of flavor-{\em weighted} energies, which are defined by $\epsilon^{\alpha}_{(t)} = {P^{\alpha}_{(t)}\,E^{\alpha}_{(t)}}$, is computed through the density matrix for {\em non-selective} measurements, $\hat{\rho}^{\prime}$, as
\begin{equation}
\sum_{\alpha = e,\mu,\tau}{\epsilon^{\alpha}_{(t)}} = \sum_{\alpha = e,\mu,\tau} {P^{\alpha}_{(t)}\,E^{\alpha}_{(t)}} =  \mbox{Tr}\{H \, \hat{\rho}^{\prime}\}.
\label{eq15B}
\end{equation}
It establishes the connection to the corresponding quantum measurement scheme \cite{Breuer} as we have described in the Appendix II.

Otherwise, the {\em total} averaged energy for a composite quantum system does not depend on the measurement scheme since it is given by
\begin{eqnarray}
E_{(t)} = \mbox{Tr}\{H \, \hat{\rho}\} &=& \sum_{\alpha = e,\mu,\tau} w_{\alpha} \, \mbox{Tr}\{H \, M^{\alpha}_{(t)}\} =  \sum_{\alpha = e,\mu,\tau} w_{\alpha} \, E^{\alpha}_{(t)},
\label{eq07}
\end{eqnarray}
from which one can notice that each energy component $E^{\alpha}_{(t)}$ for $\alpha = e,\,\mu,\,\tau$ is respectively decoupled from its corresponding statistical weight $w_{\alpha}$.
Flavor {\em averaged} energies, $E^{\alpha}_{(t)}$, are not correlated to flavor probabilities described by Eq~(\ref{eq06}) since such probabilities, $P^{\alpha}_{(t)}$, have multiple dependencies on all statistical weights, $w_{e}$, $w_{\mu}$ and $w_{\tau}$.
The inaccuracy in correlating flavor-{\em averaged} energies, $E^{\alpha}_{(t)}$, to flavor eigenstates is therefore obvious.
However, there are no ambiguities in defining the {\em total} averaged energy, $E_{(t)}$.

Besides being consistently embedded into the quantum measurement scheme \cite{Breuer}, the comparison between flavor-{\em averaged} and flavor-{\em weighted} energy definitions \cite{Bernardini} reported above allows us to disambiguate the correspondence between flavor eigenstate energies and measurement probabilities.
To clarify this point, we perform some mathematical manipulations involving the projection operators from Eq.~(\ref{eq01B}) through which one easily finds that
\begin{eqnarray}
M^{\alpha}_{(0)} \, \hat{\rho}\, M^{\alpha}_{(0)} &=&
\sum_{\beta = e,\mu,\tau} (w_{\beta} \, \mathcal{P}_{\beta\rightarrow \alpha}\bb{t})\, M^{\alpha}_{(0)}
= P^{\alpha}_{(t)}\, M^{\alpha}_{(0)},
\label{eq08}
\end{eqnarray}
where $\hat{\rho}_{\alpha} \equiv M^{\alpha}_{(0)}$, so that flavor-{\em weighted} energies can be rewritten as
\begin{eqnarray}
\epsilon^{\alpha}_{(t)} = \mbox{Tr}\{M^{\alpha}_{(0)}\,H\, M^{\alpha}_{(0)} \, \hat{\rho} \} &=& \mbox{Tr}\{H\, M^{\alpha}_{(0)} \, \hat{\rho}\, M^{\alpha}_{(0)}\}
= P^{\alpha}_{(t)}\mbox{Tr}\,\{M^{\alpha}_{(0)}\,H\} =
P^{\alpha}_{(t)}\, E^{\alpha}_{(t)}.
\label{eq09}
\end{eqnarray}
It can be promptly correlated to the previous definition through the relation
\begin{equation}
\sum_{\alpha = e,\mu,\tau} \left(\epsilon^{\alpha}_{(t)} - w_{\alpha} E^{\alpha}_{(t)}\right) = \sum_{\alpha,\beta = e,\mu,\tau}^{\alpha\neq\beta} \mbox{Tr}\{M^{\alpha}_{(0)}\,H\, M^{\beta}_{(0)} \, \hat{\rho} \},
\label{eq10A}
\end{equation}
from which, in analogy to the quantum interference phenomenon, one can depict a residual interference effect since it intrinsically brings simultaneous information of mixed flavor eigenstates.

One should notice that the time-averaged value of the above-obtained residual term is not null.
Consequently it leads to different interpretations for the mean values of flavor-{\em averaged} and flavor-{\em weighted} energies. The former one introduces a ill-defined relation between energy and probability, and the latter one provides us with the probabilistic correlation between observable energies and flavor quantum numbers (see the Appendix II for details).

From this point, one can thus turn his attention to the consequences of the above analysis in predicting the cosmological neutrino mass effective values.

The usual method for computing the energy density of neutrinos in the Universe follows from the {\em averaged} mass from Eq.~(\ref{1}) that results into
\begin{equation}
m^2_{\rm{eff}, \nu_\alpha} = \sum_i \vert U_{\alpha i} \vert^2 m_i^2,
\label{meff}
\end{equation}
and from a {\em weighted} number density distributions of neutrinos in flavor eigenstates,
\begin{equation}
d n_{\nu_i} = \sum_\alpha \vert U_{\alpha i} \vert^2 d n_{\nu_\alpha},
\label{eq22AA}
\end{equation}
so that, as reported by \cite{Fuller}, one could have
\begin{equation}
\rho^{(Std)}_E = \sum_\alpha \int ( p^2 + m^2_{\rm{eff}, \nu_\alpha} )^{1/2} d n_{\nu_\alpha}.
\label{naive}
\end{equation}
Here one has assumed that the neutrino momentum distribution for each flavor can be approximated by Fermi-Dirac distribution functions, with the number density $\nu_\alpha$'s in a momentum interval $dp$ given by \cite{Fuller}
\begin{equation}
\label{dist}
dn_{\nu_\alpha}= {{1}\over{2\pi^2}}\cdot {{p^2}\over{ e^{E_{\nu_{\alpha}}(a)/ T(a) - \eta_{\nu_\alpha}} +1}}\,dp,
\end{equation}
where it is assumed natural units given by $\hbar = c = k_B = 1$, and it is introduced the ratio of chemical potential to temperature for neutrino species $\nu_\alpha$, $\eta_{\nu_\alpha}$.

In a previous issue \cite{Fuller}, it has been supposed that to calculate the energy density of these particles in a quantum mechanically consistent way, the mass eigenstate energies should be considered in order to give
\begin{eqnarray}
\rho^{\nu}_E & = & \sum_i \int ( p^2 + m_i^2 )^{1/2} d n_{\nu_i} \nonumber \\
 & = & \sum_{i, \alpha} \int \vert U_{\alpha i} \vert^2 ( p^2 + m_i^2 )^{1/2} d n_{\nu_\alpha} .
\label{std}
\end{eqnarray}
From the analysis performed in \cite{Fuller}, the distribution functions of neutrinos in mass eigenstates would not have a Fermi-Dirac form when the degeneracy parameters were not identical for all the three active flavors.
It certainly would lead to ambiguities in the confront between the above-defined energy densities.
From the density matrix framework, the {\em total} averaged energy is unequivocally defined through Eq.(\ref{eq07}).
Otherwise, if one considers the flavor-{\em weighted} energy correlated to flavor probabilities for computing the total neutrino energy density $\rho_E$, at least the residual divergence described by Eq.~(\ref{eq10A}) would remain pertinent.
However, the subtleties circumventing the general theory of quantum measurements for composite quantum systems allow one to quantify such a residual interference contribution through the statistical weights.
We thus assume that the energy-momentum dispersion relation related to $E_{\nu}$ has to be defined through the quantum measurement scheme.
And we shall see in the following that all the ambiguities disappear in case of a maximal statistical mixture.

%%%%%%%%%%%%%%%%%%%%%%%%%%%%%%%%%%%%%%%%%%%%%%%%%%%%%%%%%%%%%%%%%%%%%%%%%%%%%%
%%%%%%%%%%%%%%%%%%%%%%%%%%%%%%%%%%%%%%%%%%%%%%%%%%%%%%%%%%%%%%%%%%%%%%%%%%%%%%
%%%%%%%%%%%%%%%%%%%%%%%%%%%%%%%%%%%%%%%%%%%%%%%%%%%%%%%%%%%%%%%%%%%%%%%%%%%%%%

We assume that each of the flavor sub-ensemble is described by a normalized state vector $\nu^{\alpha}$, with $\alpha = e, \, \mu, \, \tau$, in the underlying Hilbert space.
It is then natural to study the statistics of the complete ensemble by mixing the flavor sub-ensembles with respective weights $w_{\alpha}$.
The mixing is achieved by taking a large number $N_{\alpha}$ of systems for each flavor ensemble so that $w_{\alpha} = N_{\alpha} / (\sum{N_{\alpha}})$.
The resulting density matrix is consistent with the assumption of instantaneous and simultaneous decoupling for all three flavors (which is indeed not realistic).
The (time-independent) coefficients $w_{\alpha}$ can then be defined via $dn_{\nu_\alpha} = w_{\alpha} \, dn_{Total}$ for
some reference phase space element $dn_{Total}$.
The maximal statistical mixture results from the natural assumption that $dn_e = dn_{\mu} = dn_{\tau}$, i. e. $w_e = w_{\mu} = w_{\tau}$.
In this case, the {\em total} averaged energy is connected to a series of amazing convergent results described by
\begin{equation}
E_{(t)} = \sum_{\alpha = e,\mu,\tau} w_{\alpha} E^{\alpha}_{(t)} = \sum_{\alpha = e,\mu,\tau} P^{\alpha}_{(t)} E^{\alpha}_{(t)} = \sum_{\alpha = e,\mu,\tau} \langle \epsilon^{\alpha}\rangle = \frac{1}{3}\sum^{3}_{s=1}{E_s} = \bar{E}.
\label{eq26}
\end{equation}
where $s = 1,\,2,\,3$ are the indices for the mass eigenstates.
One then concludes that all the flavor energy definitions derived from the generalized quantum measurement framework reproduce exactly the same results for the neutrino energy density when the neutrino ensemble is a maximal statistical mixing.
Equivalently, for a $D$-dimension mass eigenstate system, the input into Eq.~(\ref{std}) used to compute the neutrino energy density, $\rho^{\nu}_E$, could be given by the average of the mass eigenstate energies, $\bar{E} = \frac{1}{D}\sum^{D}_{s=1}{E_s}$.

\begin{figure}
\epsfig{file= 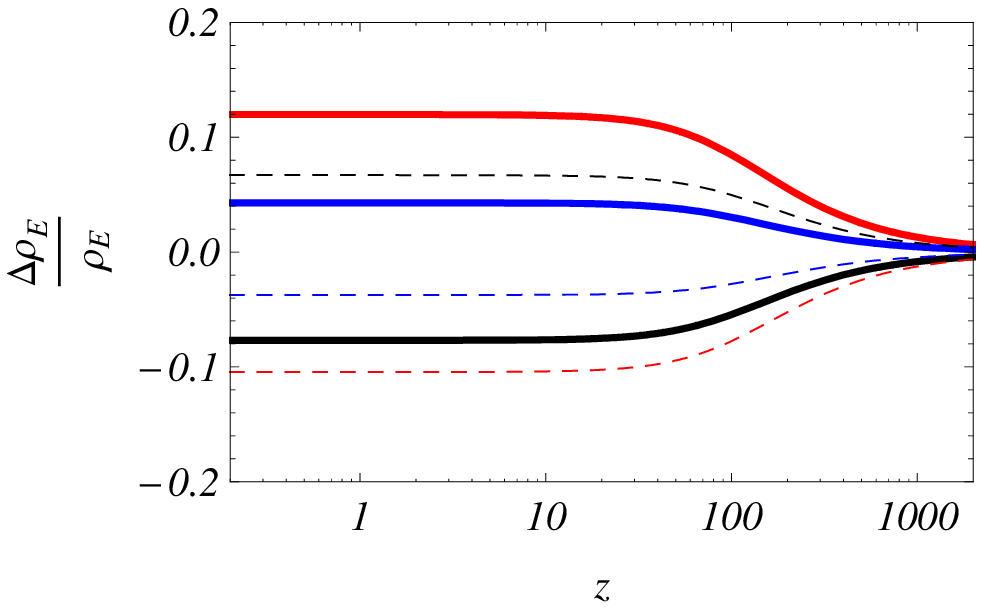, width= 11 cm}
\epsfig{file= 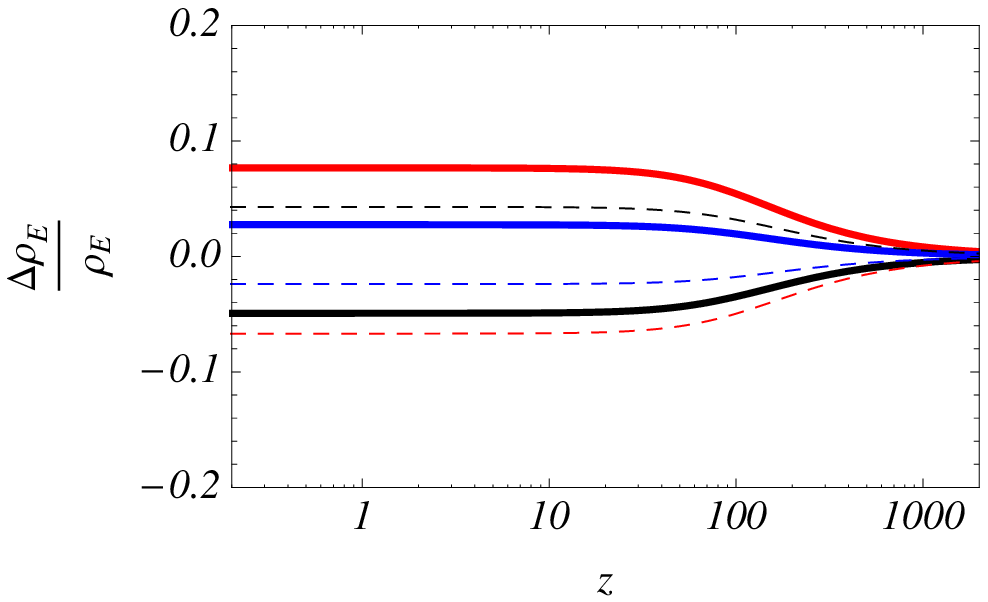, width= 11 cm}
\caption{\label{an4} Energy density deviations, $\Delta \rho_{E}/ \rho_{E}$ as a function of redshift, $z$.
Solid curves are for $m_{\1} \simeq 300 \, k_{B} T^{\nu}_{\0} = 50 ~{\rm meV}$ in the normal neutrino mass hierarchy and dashed curves are for $m_{\3} \simeq 300 \, k_{B} T^{\nu}_{\0} = 1.67 ~{\rm meV}$ in the inverted hierarchy.
We have assumed that $\Delta m^{\2}_{\rm atm} \simeq 2.4 \times 10^{-3}\, {\rm eV}^{\2}$ and $\Delta m^{\2}_{\rm \bigodot} = m^{\2}_{\1}-m^{\2}_{\2} \simeq 7.6 \times 10^{-5}\, {\rm eV}^{\2}$.
We reproduce the relative difference between the {\em total} averaged and flavor-{\em weighted} energies, $\langle E_{(t)} \rangle$ and $(\sum_{\alpha}{E^{\alpha}})$ (black lines), between the quantum mechanical mass averaged and the {\em total} averaged energies, $\bar{E}$ and $\langle E_{(t)} \rangle$ (red lines), and between the quantum mechanical mass averaged and the flavor-{\em weighted} energies, $\bar{E}$ and $(\sum_{\alpha}{E^{\alpha}})$ (blue lines).
We have assumed the current phenomenological values for the neutrino mixing angles, i. e. $\theta_{12} \approx 0.5905$, $\theta_{23} \approx \pi/4$, and $\theta_{13} = 0$.
In the first plot, the results are for a pure state of electronic neutrinos, i. e. $w = w_{e} = 1$ corresponding to the theoretical maximal bounds.
In the second plot, the results are for a realistic statistical mixing in correspondence to Eq.~(\ref{234}), with $w_{e}\sim 0.76$, and $w_{\mu} \approx w_{\tau}\sim 0.12$.}
\end{figure}

\begin{figure}
\epsfig{file= 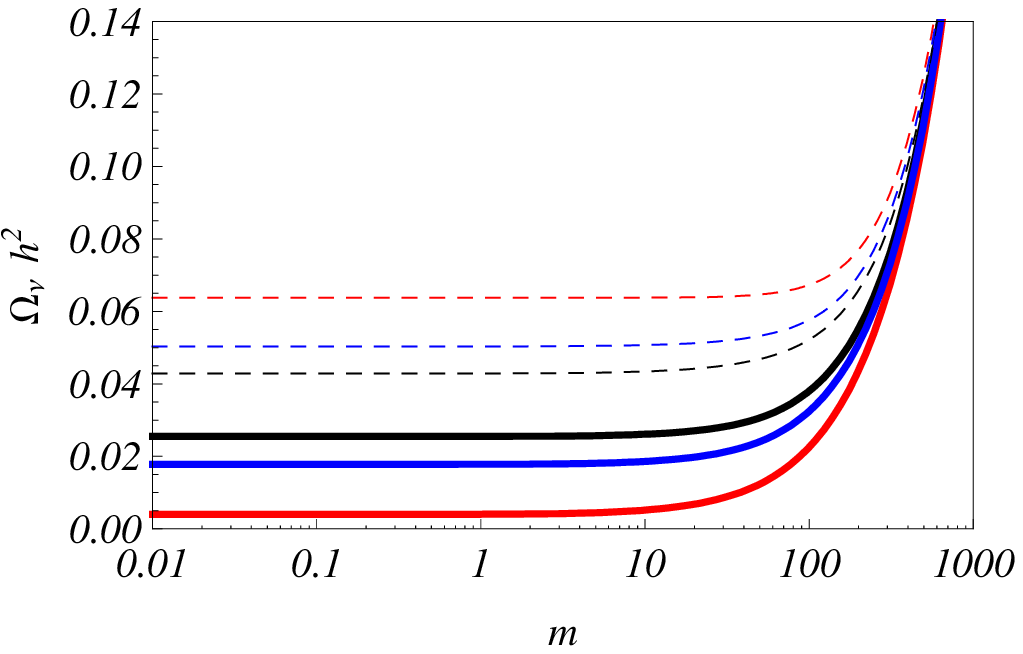, width= 11 cm}
\epsfig{file= 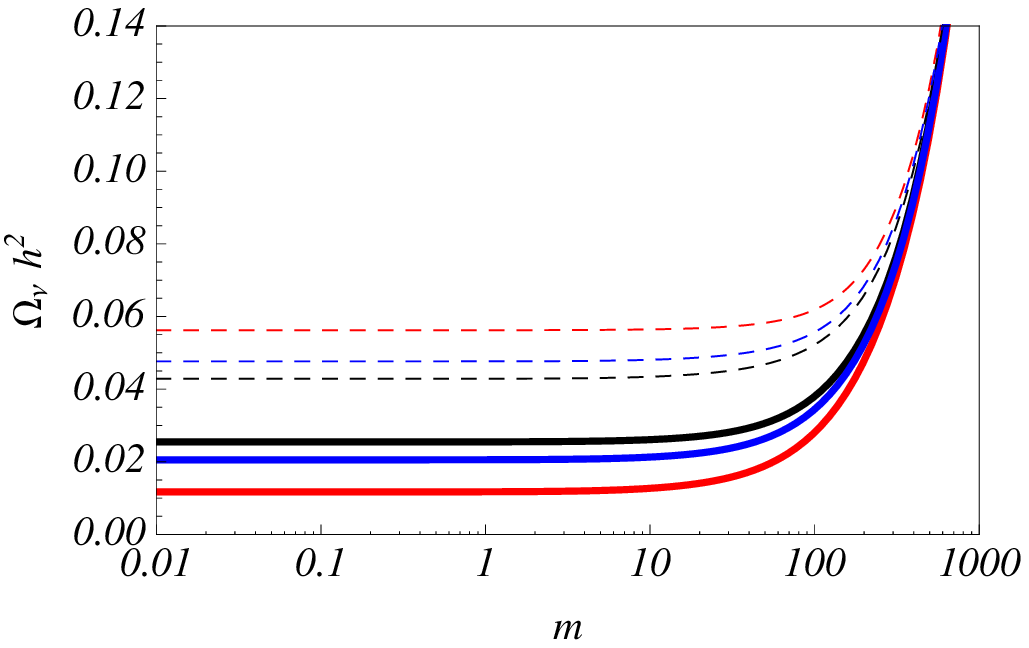, width= 11 cm}
\caption{\label{an6} Total neutrino energy density, $\Omega_{\nu} h^2$,as a function of the lightest mass eigenvalue $m$ in units of $k_{B} T^{\nu}_{\0} = 0.167 ~{\rm meV}$.
Solid curves are for the normal neutrino mass hierarchy and dashed ones for the inverted hierarchy.
Again we have assumed that $\Delta m^{\2}_{\rm atm} \simeq 2.4 \times 10^{-3}\, {\rm eV}^{\2}$ and $\Delta m^{\2}_{\rm \bigodot} = m^{\2}_{\1}-m^{\2}_{\2} \simeq 7.6 \times 10^{-5}\, {\rm eV}^{\2}$.
The line colors, the input parameters and the phenomenological assumptions are in correspondence with Fig.~\ref{an4}.
The results depicted from the first plot are for a pure state of electronic neutrinos, i. e. $w = w_{e} = 1$, and the results depicted from the second plot are for a statistical mixing in correspondence with Eq.~(\ref{234}), with $w_{e}\sim 0.76$, and $w_{\mu} \approx w_{\tau}\sim 0.12$.}
\end{figure}

\begin{figure}
\epsfig{file= 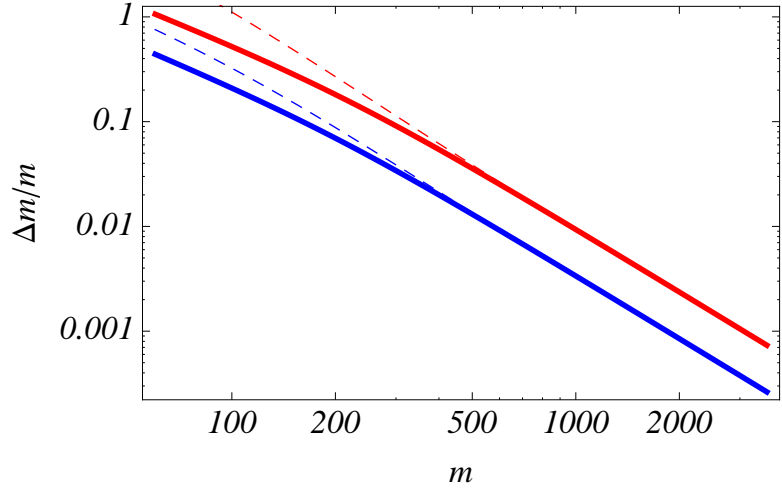, width= 11 cm}
\caption{\label{an7} Fractional error, $\Delta m/m$, for the neutrino mass value predictions as a function of the lightest mass eigenvalue, $m$, in units of $k_{B} T^{\nu}_{\0} = 0.167 ~{\rm meV}$.
The solid line is for the normal neutrino mass hierarchy and the dashed one for the inverted hierarchy (and assumptions are in correspondence with the second plot of Figs.~\ref{an4} and ~\ref{an6}).}
\end{figure}

By following the arguments of \cite{Bernardini} and observing that $H_{\0}^{-1} \sim 0.7 \times 10^{33} \,{\rm eV}^{-1}$, $\Delta m^{2} \lesssim 2.4 \times 10^{-3} \,{\rm eV}^{2}$ and $q \sim 0.167 \times 10^{-4}\, {\rm eV}$, one finds a huge oscillation number given by $\Delta E\, \tau \sim 10^{34}$, that qualifies the time-average ($\langle \rangle_{\rm time}$) procedure as a good approach for computing the explicit values of energies and probabilities.
The time-averaged flavor probabilities are easily obtained as $\langle P^{e,\mu,\tau}\rangle =  \frac{1}{3}$ and
and the corresponding time-averaged flavor-{\em weighted} energies would be given by
\begin{eqnarray}
\langle \epsilon^{e} \rangle &=& \frac{\bar{E}}{3} + \frac{1}{36}\left[ 6\, \delta_{\1\2}\, \cos\bb{2\theta_{\1\2}} \, \cos^{\2}\bb{\theta_{\1\3}} - (\delta_{\2\3} - \delta_{\3\1}) (1 - 3 \cos\bb{2\theta_{\1\3}})\right]\nonumber\\
\langle \epsilon^{\mu} \rangle &=& \frac{\bar{E}}{3} + \frac{1}{9}\left[
(\delta_{\3\1} - \delta_{\2\3})
\cos^{\2}\bb{\theta_{\1\3}}\,\sin^{\2}\bb{\theta_{\2\3}}\right.\nonumber\\
&& +
(\delta_{\1\2}-\delta_{\3\1})
\left(\sin\bb{\theta_{\1\2}}\,\cos\bb{\theta_{\2\3}}
+
\cos\bb{\theta_{\1\2}}\,\sin\bb{\theta_{\1\3}}\,\sin\bb{\theta_{\2\3}}\right)^{\2}\nonumber\\
&&\left.+
(\delta_{\2\3} - \delta_{\1\2})
\left(\cos\bb{\theta_{\1\2}}\,\cos\bb{\theta_{\2\3}}
-
\sin\bb{\theta_{\1\2}}\, \sin\bb{\theta_{\1\3}}\,\sin\bb{\theta_{\2\3}}\right)^{\2}
\right]\nonumber\\
\langle \epsilon^{\tau} \rangle &=& \frac{\bar{E}}{3} + \frac{1}{9}\left[
(\delta_{\3\1} - \delta_{\2\3})
\cos^{\2}\bb{\theta_{\1\3}}\,\cos^{\2}\bb{\theta_{\2\3}}\right.\nonumber\\
&& -
(\delta_{\1\2}-\delta_{\3\1})
\left(\cos\bb{\theta_{\1\2}}\, \sin\bb{\theta_{\2\3}} + \sin\bb{\theta_{\1\2}}\,\sin\bb{\theta_{\1\3}}\,\cos\bb{\theta_{\2\3}}\right)^{\2}\nonumber\\
&&\left.-
(\delta_{\2\3} - \delta_{\1\2})
\left(\sin\bb{\theta_{\1\2}}\,\sin\bb{\theta_{\2\3}} - \cos\bb{\theta_{\1\2}}\,\sin\bb{\theta_{\1\3}}\,\cos\bb{\theta_{\2\3}}\right)^{\2}
\right]
\label{eq25}
\end{eqnarray}
where $\delta_{ij} = E_{i} - E_{j}$ correspond to mass eigenstate energy differences.
The above results can be summed up in order to verify the Eq.~(\ref{eq26}) and the quantum mechanical definition from Eq.~(\ref{std}) which sets the cosmological neutrino energy density dependence on the sum of the mass eigenstate energy eigenvalues through its relation with $\bar{E}$.
In case of pure states and non-maximal statistical mixings, flavor-{\em weighted} energies lead to different predictions for $\rho_E$, as one can notice through the results for the energy density deviations from Fig.~\ref{an4}.
Reproducing qualitatively the effects obtained from previous issues \cite{Fuller, Bernardini}, Fig.~\ref{an4} shows that, at early times, neutrino momenta are large enough that all the mass eigenstates are ultrarelativstic and masses have a tiny influence on the total neutrino energy density.
The ultrarelativistic regime naturally suppress any eventual divergence from the naive effective mass approach.
Complementing the results from Fig.~\ref{an4}, the total neutrino energy density, $\Omega_{\nu} h^2$, in correspondence with the lightest mass eigenvalue $m$ can be depicted from Fig.~\ref{an6}.
And finally, the fractional error, $\Delta m/m$, for the neutrino mass value predictions as function of the lightest mass eigenvalue, $m$, for normal and inverted hierarchies are depicted from Fig.~\ref{an7}.

Despite the evident divergencies between the results of measurement schemes depicted by Figs.~\ref{an4}-\ref{an7} for non-maximal statistical mixings, we reinforce that, in case of a maximal statistical mixing (i. e. when $\delta w = 0$), all the measurement schemes reproduce the predictions from Eq.(\ref{std}).

In case of a maximal statistical mixing (i. e. in case of averaging out the off diagonal terms of the density matrix) all the definitions of energy that we have explored lead to the same results for the eventually measured quantity. 
Since the last (elastic) scattering surface of cosmological neutrinos corresponds to electron-neutrino interactions (a selective measurement procedure), one could accept that the neutrino free-streaming evolution departs from a non-maximal statistical mixing.
In this case, different energy definitions should give different outputs.

In fact, its is partly correct that $e$-flavor neutrinos are the last to interact.
The other flavors decouple from the plasma slightly earlier due to the strictly neutral current interactions of the $\mu$ and $\tau$ with the electron/positron plasma.
The slightly coupled $e$-flavor neutrinos are known to suffer a slight increasing of temperature \cite{Dodd} due to the annihilation of electrons and positrons at that point and, when including flavor oscillations \cite{Mangano}.
In the standard picture of neutrino decoupling in the early universe, the three neutrino species ($e$, $\mu$, $\tau$) are kept in thermal equilibrium with the radiation plasma through the elastic scattering process with electrons(positrons).
As a quantum mixing, neutrinos ($e$, $\mu$, $\tau$) coexisting approximately with the {\em same averaged temperature}, i. e. neutrinos corresponding to the same element of the phase space (when it is constrained by some momentum/temperature distribution), reach the thermal equilibrium through a {\em measurement} scheme produced by the elastic scattering.
The proportion between the corresponding cross sections, $\sigma_{\nu}$, is given by
\begin{equation}
\sigma_e \,:\,\sigma_{\mu}\,:\, \sigma_{\tau} \quad \Leftrightarrow \quad 1\,:\,0.16\,:\,0.16.
\end{equation}
After scattering ends up, one should have an averaged statistical ensemble described by
\begin{equation}
1\,:\,0.16\,:\,0.16 \quad \Leftrightarrow \quad w_e \,:\,w_{\mu}\,:\,w_{\tau},
\label{234}
\end{equation}
and for the values corresponding to the rapport from Eq.~(\ref{234}), one should have $w_{e}\sim 0.76$, and $w_{\mu} \approx w_{\tau}\sim 0.12$.
The plots for such a more realistic case shows that the energy density deviations are slightly suppressed when compared to the events for an electronic pure state.
However, the deviations are still relevant as one can depict from the second plot of Fig.~\ref{an4}. 
The corresponding modifications are reproduced by the second plot of Fig.~\ref{an6}, where the energy density deviations are explicitly computed.

The results depicted from Figs.~\ref{an4} and ~\ref{an6} does not change the significance and the magnitude of the finite-temperature electromagnetic corrections to the energy density of the $\gamma e^+ e^-$ radiation plasma \cite{Dodd,142,143} or of the finite temperature QCD corrections \cite{Dolgov02}.
They are of the same order of magnitude of flavor mixing corrections upon the averaged temperature of decoupling for different neutrino species, which also depend on the mixing parameters \cite{Dolgov02}.

The energy dependence of an ensemble of neutrino flavors on the statistical weights can be determined and the role of different flavor energy definitions in obtaining the expressions for cosmological neutrino masses can therefore be discriminated.
At the viewpoint of the theory of quantum measurement, the concept of flavor-{\em weighted} energies correlated to flavor probabilities is indeed relevant in resolving the ambiguities and misunderstandings that arise when flavor-{\em averaged} energies are defined.
We have found that the most appropriate relation between the cosmological neutrino background energy density and neutrino mass values, in case of a maximal statistical mixing, is given by the same result obtained from the single-particle quantum mechanics.

From the theoretical perspective, our analysis is at least relevant in defining the correct expression for the effective mass value of neutrinos used in the confront with the cosmological data.
From the phenomenological point of view, some of the most recent neutrino mass claims provide us with some effective mass values given in terms of
$\sum\, m_{\alpha} < 0.36\,eV$ \cite{Putter} and $\sum\,m_{\alpha} \sim 0.1 - 0.6\,eV$
\cite{Aba}.
The mass fractional error, $\Delta m/m$,  that we have addressed through the results from Fig.~\ref{an7} are correspondingly given by,
\begin{eqnarray}
m = \,0.10\, eV~~\Rightarrow&\Delta m/m = \,0.009\, (0.027)&\qquad\mbox{blue (red) lines;} \nonumber\\
m = \,0.36\, eV~~\Rightarrow&\Delta m/m = \,0.0008\, (0.0020)&\qquad\mbox{blue (red) lines;} \nonumber\\
m = \,0.60\, eV~~\Rightarrow&\Delta m/m = \,0.0003\, (0.0008)&\qquad\mbox{blue (red) lines;}
\end{eqnarray}
for both hierarchies.
Notice that at mass scales $\sum\,m_{\alpha} > 0.1\,eV$, normal and inverted hierarchies lead to corrections of the same magnitude. As one can notice, for the non-relativistic regime the corrections does not reach 1\% of the phenomenological values. The corrections increase exponentially for smaller values of the mass scale.

The consistency of our approach with previous quantum mechanics predictions and its theoretical support provided by the fundamentals of the generalized theory of quantum measurements have shown that the correct interpretation of flavor associated energies demands for a statistical description through density matrices.
It is important to emphasize that without determining the measurement procedure that should be connected to the phenomenology, our analysis does not provide the definitive answer to which energy should be used to set the cosmological neutrino mass bounds.
Therefore, at first glance, our manuscript concerns with introducing the problem of defining observable quantities from a single particle quantum system approach and comparing them with those obtained from a composite quantum system framework.

To summarize, our aim was to establish a debugged correlation between {\em averaged}/{\em weighted} energy definitions and {\em selective}/{\em non-selective} quantum measurements \cite{Bernardini}.
Generically speaking, it arises when one considers either an ensemble of systems, or a composite quantum system defined when its preparation history is uncertain and one does not know whether it is a pure quantum state or a statistical mixture.
An ensemble of neutrino flavor eigenstates in the cosmological scenario was the example that we have discussed here.

Finally, as it has been extensively discussed, the closure fraction of cold dark matter at present substantially modifies the matter power spectrum, even for neutrinos behaving like hot dark matter at higher redshifts \cite{Bernardini02}.
It follows that the amount of cold dark matter at earlier epochs should be reduced, suppressing the formation of large scale structures, when it is compared to a situation without massive neutrinos.
Such a suppression is attenuated at intermediate scales \cite{Bernardini02} if neutrinos are treated as hot dark matter.
Using the correct expression for neutrino masses can slightly modify the characterization of neutrinos as cold or hot dark matter.
It is therefore a relevant aspect that has to be included in the procedures for determining the fraction of the neutrino energy density at late times.
Our results provide conditions for understanding the background quantum mechanics of such procedures that lead to more accurate phenomenological predictions for the analysis where cosmological neutrino masses are considered.

\begin{acknowledgments}
The authors would like to thank for the financial support from the Brazilian Agencies FAPESP (grants 2008/50671-0, 2009/50959-7 and 2010/03561-0) and CNPq (grant 300233/2010-8).
\end{acknowledgments}

\section*{Appendix I - Flavor associated energies for a two level system}

The time evolution of a quantum system of well-defined flavor quantum numbers described by the state vectors $\nu^{e}$ and $\nu^{\mu}$ respectively related to electron and muon neutrinos is given by
\begin{equation}
\left(\begin{array}{l}\nu^e_{(t)} \\ \nu^{\mu}_{(t)} \end{array}\right) =
U \, \left(\begin{array}{ll} e^{-i\, E_{\1} t} & 0 \\ 0  &  e^{-i\, E_{\2} t} \end{array}\right)\,
\left(\begin{array}{l}\nu_1 \\ \nu_2 \end{array}\right) =
U \, \left(\begin{array}{ll} e^{-i\, E_{\1} t} & 0 \\ 0  &  e^{-i\, E_{\2} t} \end{array}\right)
\, U^{\dagger} \left(\begin{array}{l} \nu^e_{(0)} \\ \nu^{\mu}_{(0)}  \end{array}\right),
\label{101A}
\end{equation}
where $\nu_{\1}$ and $\nu_{\2}$ are the mass eigenstates with well-defined energies, $E_{s} = \sqrt{p^{\2} + m^{\2}_{s}}$, with $s = 1,\, 2$, and the matrix $U$ parameterizes the mixing relation as
\begin{equation}
\left(\begin{array}{l}\nu^e_{(0)} \\ \nu^{\mu}_{(0)} \end{array}\right) =
U \, \left(\begin{array}{l}\nu_1 \\ \nu_2 \end{array}\right) =
\left(\begin{array}{rr} \cos\bb{\theta} & \sin\bb{\theta} \\ -\sin\bb{\theta}  & \cos\bb{\theta} \end{array}\right)
\left(\begin{array}{l}\nu_1 \\ \nu_2 \end{array}\right),
\label{req00A}
\end{equation}
where $\theta$ is the mixing angle.
Since the Hamiltonian of the system in the mass eigenstate basis can be extracted from Eq.~(\ref{req00A}) as $H = Diag\{E_{\1},\, E_{\2}\}$, the flavor projection operators can be easily defined as
\begin{equation}
M^{e}_{(t)} = |\nu^{e}_{(t)} \rangle \langle \nu^{e}_{(t)}| =
\left[
\begin{array}{ll}
\cos^{\2}\bb{\theta} & \sin\bb{\theta} \,\cos\bb{\theta} \,e^{-i\,\Delta E\, t}\\
\sin\bb{\theta}\, \cos\bb{\theta} \, e^{i\,\Delta E\, t} &  \sin^{\2}\bb{\theta}
\end{array}
\right]
\label{req01A}
\end{equation}
and
\begin{equation}
M^{\mu}_{(t)} = |\nu^{\mu}_{(t)} \rangle \langle \nu^{\mu}_{(t)}|  =
\left[
\begin{array}{ll}
\sin^{\2}\bb{\theta} & -\sin\bb{\theta} \,\cos\bb{\theta} \,e^{-i\,\Delta E\, t}\\
-\sin\bb{\theta}\, \cos\bb{\theta} \, e^{i\,\Delta E\, t} &  \cos^{\2}\bb{\theta} \end{array}
\right]
\label{req01B}
\end{equation}
where $\Delta E = E_{\1} - E_{\2}$ and it can be verified that $M^{e}_{(t)} + M^{\mu}_{(t)}  = \mathbf{1}$.

Thus the temporal evolution of a flavor eigenstate can be described by
\begin{equation}
|\nu^{e,\mu}_{(t)} \rangle = (M^{e}_{(0)} + M^{\mu}_{(0)}) |\nu^{e,\mu}_{(t)} \rangle =
 \langle \nu^{e}_{(0)} |\nu^{e,\mu}_{(t)} \rangle \, |\nu^{e}_{(0)} \rangle
+
\langle \nu^{\mu}_{(0)} |\nu^{e,\mu}_{(t)} \rangle \, |\nu^{\mu}_{(0)} \rangle,
\label{req01C}
\end{equation}
and the supposedly relevant measurable quantities, or {\em observables}, of the closed quantum system can be summarized by the the flavor-{\em averaged} energies,
\begin{eqnarray}
E^{e,\mu}_{(t)} &=&  \langle \nu^{e,\mu}_{(t)} |H| \nu^{e,\mu}_{(t)}\rangle,
\label{req02}
\end{eqnarray}
that result in time-independent quantities,
\begin{eqnarray}
E^{e}_{(t)} &=& E^{e}_{(0)} = \bar{E} + (1/2) \Delta E\, \cos\bb{2\theta},\nonumber\\
E^{\mu}_{(t)} &=& E^{\mu}_{(0)} = \bar{E} - (1/2) \Delta E\, \cos\bb{2\theta},
\label{req02A}
\end{eqnarray}
with $\bar{E} = (1/2)(E_{\1} + E_{\2})$, and by the time-oscillating flavor probabilities,
\begin{eqnarray}
\mathcal{P}_{\alpha\rightarrow\beta} \bb{t} &=& Tr\{M^{\beta}_{(0)}\, M^{\alpha}_{(t)}\}, ~~~~\alpha,\,\beta = e,\, \mu,
\label{req02BB}
\end{eqnarray}
that result in
\begin{eqnarray}
\mathcal{P}_{e\rightarrow e} \bb{t} &=& P_{\mu\rightarrow\mu} \bb{t} = |\langle \nu^{e}_{(\0)}|\nu^{e}_{(t)}\rangle|^{\2} = |\langle \nu^{\mu}_{(\0)}|\nu^{\mu}_{(t)}\rangle|^{\2}  = 1 - \sin^{\2}\bb{2\theta} \,\sin^{\2}\left(\frac{\Delta E}{2}\, t\right),
\label{req02B}
\end{eqnarray}
and
\begin{eqnarray}
\mathcal{P}_{e\rightarrow \mu}\bb{t} &=& P_{\mu \rightarrow e} \bb{t}= |\langle \nu^{\mu}_{(\0)}|\nu^{e}_{(t)}\rangle|^{\2}  = |\langle \nu^{e}_{(\0)}|\nu^{\mu}_{(t)}\rangle|^{\2} = \sin^{\2}\bb{2\theta} \,\sin^{\2}\left(\frac{\Delta E}{2}\, t\right),
\label{req02C}
\end{eqnarray}
that are interpreted as the probabilities of $e(\mu)$-flavor states produced at time $t_{\0}$ be measured as $e(\mu)$-flavor states or be converted into $\mu (e)$-flavor states after a time interval $t - t_{\0} \sim t - 0 \sim t$.

Now let us suppose that the density matrix of a composite quantum system of two neutrino flavor states is given by
\begin{equation}
\hat{\rho}\bb{t} \equiv \hat{\rho} = w_e \, M^{e}_{(t)} + w_{\mu} \, M^{\mu}_{(t)},
\label{req05}
\end{equation}
with $w_{e} + w_{\mu} = 1$.
One easily finds that the re-defined probabilities of measuring the electron and muon flavor eigenstates at time $t$ are given by
\begin{eqnarray}
P^{e}_{(t)}   = \mbox{Tr}\{M^{e}_{(0)}\, \hat{\rho}\} &=& w_e \, \mbox{Tr}\{M^{e}_{(0)}\,M^{e}_{(t)}\} + w_{\mu} \, \mbox{Tr}\{M^{e}_{(0)}\,M^{\mu}_{(t)}\} = \nonumber\\
                        &=& w_e \, \mathcal{P}_{e\rightarrow e}\bb{t} + w_{\mu} \, \mathcal{P}_{\mu\rightarrow e}\bb{t},\\
P^{\mu}_{(t)} = \mbox{Tr}\{M^{\mu}_{(0)}\, \hat{\rho}\} &=& w_e \, \mbox{Tr}\{M^{\mu}_{(0)}\,M^{e}_{(t)}\} + w_{\mu} \, \mbox{Tr}\{M^{\mu}_{(0)}\,M^{\mu}_{(t)}\} = \nonumber\\
                        &=& w_e \, \mathcal{P}_{e\rightarrow \mu}\bb{t} + w_{\mu} \, \mathcal{P}_{\mu\rightarrow \mu}\bb{t},
\label{req06}
\end{eqnarray}
where we have used the results from Eqs.~(\ref{req02B}-\ref{req02C}).
One also easily notices that
\begin{eqnarray}
P^{e}_{(t)} +  P^{\mu}_{(t)} &=&
w_{e} (\mathcal{P}_{e\rightarrow e}\bb{t} + \mathcal{P}_{e\rightarrow \mu}\bb{t})
+
w_\mu (\mathcal{P}_{\mu\rightarrow e}\bb{t} + \mathcal{P}_{\mu\rightarrow \mu}\bb{t})
= w_{e} + w_\mu = 1
\label{req06B}
\end{eqnarray}
and that the properties of a statistical mixture are immediate.
It leads to a reinterpretation of the energy related to each flavor quantum number.

The standard {\em total} averaged energy for a composite quantum system is defined through the density matrix as
\begin{eqnarray}
E_{(t)} = \mbox{Tr}\{H \, \hat{\rho}\} &=& w_{e} \, \mbox{Tr}\{H \, M^{e}_{(t)}\} + w_{\mu} \, \mbox{Tr}\{H \, M^{\mu}_{(t)}\}\nonumber\\
                            &=&   w_{e} \, E^{e}_{(t)} + w_{\mu} \, E^{\mu}_{(t)},
\label{req07}
\end{eqnarray}
from which one can notice the explicit dependence on the flavor-{\em averaged} energies, $E^{e,\mu}_{(t)}$, recovered from Eq.~(\ref{req02}).
In this context $E^{e}_{(t)}$ and $E^{\mu}_{(t)}$ are respectively decoupled from the statistical weights $w_{\mu}$ and $w_{e}$.
It just ratifies our previous arguments that such flavor energies are noway correlated with the flavor probabilities from Eq~(\ref{req07}), $P^{e}_{(t)}$ and $P^{\mu}_{(t)}$ since both of them depend simultaneously on both statistical weights, $w_{\mu}$ and $w_{e}$.
Thus the arguments that assert the ambiguity and the insufficiency in defining the flavor eigenstate averaged energies through $E^{e,\mu}_{(t)}$ are maintained.

After simple mathematical manipulations involving the definitions from Eq.~(\ref{req01B}) and the probabilities from Eq.~(\ref{req06}), one easily finds that
\begin{eqnarray}
M^{\mu}_{(0)} \, \hat{\rho}\, M^{\mu}_{(0)} &=& (w_e \, \mathcal{P}_{e\rightarrow \mu} \bb{t}+ w_{\mu} \, \mathcal{P}_{\mu\rightarrow \mu}\bb{t}) M^{\mu}_{(0)} = P^{\mu}_{(t)}\, M^{\mu}_{(0)},\nonumber\\
M^{e}_{(0)} \, \hat{\rho}\, M^{e}_{(0)}     &=& (w_e \, \mathcal{P}_{e\rightarrow e}\bb{t} + w_{\mu} \, \mathcal{P}_{\mu\rightarrow e}\bb{t}) M^{e}_{(0)} = P^{e}_{(t)}\, M^{e}_{(0)},
\label{req08}
\end{eqnarray}
Observing the cyclic properties of the trace, the flavor-{\em weighted} energies can be defined as
\begin{eqnarray}
\epsilon^{e,\mu}_{(t)} = \mbox{Tr}\{M^{e,\mu}_{(0)}\,H\, M^{e,\mu}_{(0)} \, \hat{\rho} \} &=& \mbox{Tr}\{H\, M^{e,\mu}_{(0)} \, \hat{\rho}\, M^{e,\mu}_{(0)}\}
= P^{e,\mu}_{(t)}\mbox{Tr}\,\{M^{e,\mu}_{(0)}\,H\} =
P^{e,\mu}_{(t)}\, E^{e,\mu}_{(0)},
\label{req09}
\end{eqnarray}
which can be promptly compared with the previous definition through the relation
\begin{equation}
\frac{|\epsilon^{e,\mu}_{(t)} - w_{e,\mu} E^{e,\mu}_{(0)}|}{E^{e,\mu}_{(0)}} = |w_{e} - w_{\mu}| \sin^{\2}\bb{2\theta}
\, \sin^{\2}\left(\frac{\Delta E}{2}\,t\right).
\label{req10A}
\end{equation}

As described above, flavor-{\em weighted} energies establish a unique correspondence between flavor eigenstate energies and the statistical definitions of probabilities, $P^{e,\mu}_{(t)}$.
It can not be identified through the definition of flavor-{\em averaged} energies.

\section*{Appendix II - The von-Neumann entropy and quantum measurements}

The connection to the quantum measurement theory arises from assuming the von-Neumann entropy as an important entropy functional defined in terms of the density matrix by
\begin{equation}
S\bb{\hat{\rho}} = - \mbox{Tr}\{\hat{\rho}\bb{t}\, \ln\bb{\hat{\rho}\bb{t}}\},
\label{eq26Z}
\end{equation}
where we have set the multiplicative Boltzmann constant, $k_{B}$, equal to unity.
The entropy $S\bb{\hat{\rho}}$ quantifies the departure of a composite quantum system from a pure state by measuring its time-evolved degree of mixture.
The quantum measurements can also induce modifications on the the von-Neumann entropy of the system.
The entropy changes due to a {\em non-selective} measurement scheme described by {\em operations} parameterized by the projection operators $M^{\alpha}_{(0)}$ are given by \cite{Breuer}
\begin{equation}
\Delta S = S\bb{\hat{\rho}^{\prime}} - S\bb{\hat{\rho}} \geq 0,
\label{eq26A}
\end{equation}
where $\alpha$ is the relevant quantum number and $\hat{\rho}^{\prime} = \sum_{\alpha} {P^{\alpha}_{(t)} \hat{\rho}_{\alpha}}$ so that
\begin{equation}
S\bb{\hat{\rho}^{\prime}} =
S\left(\sum_{\alpha} {P^{\alpha}_{(t)} \hat{\rho}_{\alpha}}\right).
\label{eq26B}
\end{equation}
Since $\Delta S \geq 0$, the {\em non-selective} quantum measurement never decreases the von-Neumann entropy.
To quantify the relation between {\em selective} and {\em non-selective} levels of measurement, the mixing entropy described by
\begin{equation}
\delta S =
S\left(\sum_{\alpha} {P^{\alpha}_{(t)} \hat{\rho}_{\alpha}}\right) - \sum_{\alpha} {P^{\alpha}_{(t)} S\left(\hat{\rho}_{\alpha}\right)}.
\label{eq26C}
\end{equation}
gives the difference between the entropy of a system projected by a {\em non-selective} quantum measurement, $S\left(\sum_{\alpha} {P^{\alpha}_{(t)} \hat{\rho}_{\alpha}}\right)$, and the average of the entropies of the sub-ensembles $\hat{\rho}_{\alpha}$, described by $M^{\alpha}_{(0)}$.
For the {\em selective} measurement scheme, with $M^{\alpha}_{(0)}$ denoting the creation of a single-flavor (pure) state, the mixing entropy is reduced to
\begin{equation}
\delta S = S\left(\sum_{\alpha} {P^{\alpha}_{(t)} \hat{\rho}_{\alpha}}\right),
\label{eq26D}
\end{equation}
since $S\left(\hat{\rho}_{\alpha}\right) = 0$.
The {\em non-selective} measurement described by observables like flavor-{\em weighted} energies always modifies the von-Neumann entropy by $\Delta S$ (c. f. Eq.~(\ref{eq26A})).
In the particular case of an ensemble described as a pure state, the {\em selective} measurement of flavor-{\em averaged} energies which are not expressed in terms of flavor conversion probabilities does not modify the von-Neumann entropy.
From the theoretical perspective, the von-Neumann entropy is therefore an auxiliary variable in distinguishing the measurement procedure and in classifying the measurement interventions as {\em selective} and {\em non-selective} ones, which however, as noticed above, are not complementary concepts.

All the above defined entropies satisfy some set of inequalities \cite{Breuer} which have been extensively used in different forms in the framework of quantum information theory and quantum entanglement.

\end{document}